\documentclass[aps,prev,twocolumn,preprintnumbers,floatfix,nofootinbib]{revtex4-1}
\pdfoutput=1

\usepackage{mathrsfs, amsmath, amsthm, amssymb, color, epstopdf, verbatim, hyperref, enumerate,graphicx}

\begin{document}

\title{GUT monopoles, the Witten effect and QCD axion phenomenology}
\author{Nick Houston$\,{}^a$}
\email{nhouston@itp.ac.cn}
\author{Tianjun Li$\,{}^{a,b}$}
\email{tli@itp.ac.cn}
\affiliation{${}^a$ CAS Key Laboratory of Theoretical Physics, Institute of Theoretical Physics, Chinese Academy of Sciences, Beijing 100190, China\\
				${}^b$ School of Physical Sciences, University of Chinese Academy of Sciences, No. 19A Yuquan Road, Beijing 100049, China}

\begin{abstract}
Taken individually, magnetic monopoles and axions are both well-motivated aspects of physics beyond the Standard Model.
We demonstrate that by virtue of the Witten effect, their interplay is furthermore nontrivial; monopoles break the corresponding axial symmetries explicitly and result in an axion potential lacking the usual instanton-derived exponential suppression factor.
As axions are distinguished by their extremely weak interactions, any regime where their effects are enhanced may be of interest.
As an application of these findings, we demonstrate that a phenomenologically acceptable population of grand-unification monopoles in the early Universe can efficiently suppress both the QCD axion dark matter abundance and CMB isocurvature contribution.
\end{abstract}

\maketitle

\section{Introduction}

Magnetic monopoles and axions are arguably two of the most plausible facets of possible physics beyond the Standard Model.
The former arise naturally from the spontaneous breaking of non-Abelian gauge symmetries \cite{Polyakov:1974ek, tHooft:1974kcl}, whilst the latter provide the most credible solution to the strong CP problem in QCD \cite{Peccei:1977hh}, and are more generally a standard feature of string compactifications \cite{Svrcek:2006yi}. 

Whilst their individual phenomenology is undoubtedly rich and varied, we will argue in the following that the interplay of these phenomena is furthermore nontrivial and may be of phenomenological consequence.

The basis of this argument is the Witten effect: in the presence of a $\theta$-term, `t Hooft-Polyakov monopoles become dyons carrying an electric charge
\begin{align}
	Q_E=ne+\frac{e\,\theta}{2\pi}\,,\quad 
	n\in\mathbb{Z}\,,
	\label{Witten effect}
\end{align}
where $e$ is the usual electromagnetic coupling \cite{Witten:1979ey}.
The surprising dependence of this charge on $\theta$ evidently suggests that axions, which play the role of a dynamical $\theta$ angle, are uniquely placed to interact with monopoles.

Indeed, in a physical basis the observable angle is 
\begin{align}
	\bar\theta=\theta+\mathrm{Arg\,Det}M^{ij}\,,\quad
	\mathcal{L}_{mass}=\overline q_{iL}M^{ij}q_{jR}+h.c.\,,
\end{align}
which is evidently non-invariant under axial, such as Peccei-Quinn (PQ), transformations, where
\begin{align}
	q_f\to e^{i\gamma_5\epsilon} q_f\,.
\end{align}
Since the action of these symmetries is to shift $\bar\theta$, there are a few observations we can make.
\begin{itemize}
	\item In a monopole background axial symmetries must be directly broken, as the corresponding transformations now alter physical quantities such as $Q_E$.
	\item The resulting $\bar\theta$ dependence will be leading order in the semi-classical approximation.
	In the ordinary QCD vacuum it carries an instanton-relation exponential suppression factor.
\end{itemize}

Since the as-of-yet unobserved axions are distinguished by their feeble interactions, typically suppressed by powers of anything up to the Planck scale, any regime where their effects can be enhanced is naturally of interest.
Of course, as magnetic monopoles appear to be in short supply in our visible Universe \cite{Burdin:2014xma}, their contribution in this sense may seem to be irrelevant.

In the early Universe however, we can expect the monopole abundance to be far more significant.
Monopole-induced effects may then offer some needed relief for the various cosmological issues associated to long-lived, light scalar fields such as axions.

In particular, if the PQ symmetry breaking scale is above that of the inflationary epoch, we can face the undesirable overproduction of both axion dark matter and isocurvature perturbations in the CMB.
If it is much lower, one must instead deal with the problems associated to the corresponding topological defects in our visible Universe \cite{Marsh:2015xka}.

In exploring these points we establish the following.
\begin{itemize}
	\item In a monopole background PQ scalars develop a non-vanishing vacuum expectation value, directly breaking the corresponding symmetry and generating an axion potential proportional to the monopole number density.
	\item By then inducing an axion mass prior to the QCD phase transition, Grand Unified Theory (GUT) monopoles can suppress both the contributions of the vacuum misalignment mechanism to the  axion dark matter abundance, and the inflationary fluctuations of the axion field which lead to CMB isocurvature.
\end{itemize}
This latter point builds directly upon the efforts of \cite{Kawasaki:2017xwt,Kawasaki:2015lpf,Nomura:2015xil}, where similar effects were seen in simple models of axion electrodynamics with hidden-sector monopoles.
However by generalising to the full QFT context we find these effects are enhanced, such that no hidden sector is required.

\section{Prior work}

As noted, we are not the first to observe the possibly interesting role of monopoles in axion physics. 
There have for example been a number of studies of the effects of hidden sector monopoles on the axion abundance and isocurvature contributions \cite{Kawasaki:2017xwt,Kawasaki:2015lpf,Nomura:2015xil}, all drawing upon the original quantisation of axion electrodynamics in a monopole background performed in \cite{Fischler:1983sc}.

The foundation of these efforts was the suggestion that whilst the electromagnetic theta angle $\left\langle\theta_{EM}\right\rangle$ should typically be non-zero, it may be energetically favourable for the axion potential to locally modify so that this quantity vanishes in the vicinity of a monopole, owing to the associated electrostatic energy cost of the dyon charge induced via the Witten effect.
This results in a proximity-dependent effective axion mass.

Unfortunately, since the electrostatic forces responsible for this phenomenon can be efficiently screened by light fermions, it was found that unless the monopoles in question belong to a hidden sector, their effects were largely negligible.

However, in the circumstances it is not entirely clear if these efforts fully capture the interplay of monopoles and axions.
Perhaps critically, QCD effects, ordinarily responsible in large part for the properties of the axion, are absent.
Furthermore, the approach taken offers no illumination of the role of monopoles in breaking PQ symmetry directly, the anomalous nature of the resultant axion mass, or more generally the realisation of these effects within the full gauge theory context. 

This being the case we will examine these effects within the full QFT framework, with the hopes of then clarifying and extending previous analyses.
For concreteness we will focus on the QCD axion, although the resulting conclusions should have general applicability.

\section{Monopole/axion interactions}

Before delving into technicalities, it may be illuminating to outline some qualitative expectations for the parameters of interest.
\subsection{Axion mass}
It is key to remember that as a pseudo-Goldstone mode, the mass of the axion derives entirely from the effects of axial anomalies.
In a monopole background we can quantify this via the anomalous non-conservation of the axial current
\begin{align}
	\partial_\mu J^\mu_5=\frac{g^2}{8\pi^2} \mathrm{Tr}\left(F_{\mu\nu}\tilde  F^{\mu\nu}\right)\simeq\frac{M}{|x|^3}\,,
	\label{axial anomaly}
\end{align}
where $|x|$ is the radial distance from the monopole and $M$ is some unknown mass scale.
The latter equivalence follows on dimensional grounds, because integrating the monopole topological density $F\tilde F$ over all space must result in a mass-dimension one constant  \cite{Marciano:1976as}
\footnote{Strictly speaking this equivalence is valid for dyons rather than monopoles, however prior to the PQ mechanism occurring $\theta\neq 0$ and so the logic is sound.
Also, in the GUT context we expect multiple $\theta$ angles which cannot typically be cancelled simultaneously, and so the fundamental `monopole' will in any case be likely dyonic.
In any case, typical monopole-induced anomalous processes also involve excitation of the dyon mode \cite{Nair:1983ps}.}.

Evidently in a single monopole background we can expect a proximity-dependent mass, as the anomalous effects are correspondingly enhanced as we move closer and closer to the monopole core.
In the more realistic scenario of a monopole/antimonopole plasma, we can use the monopole number density to provide an effective average mass, based on the typical inter-monopole separation distance.

Prior to the QCD phase transition, this will yield the only contribution to the axion mass.
Below this temperature, the ordinary QCD axial anomaly will provide the dominant contribution at low energies and/or monopole densities.

\subsection{Decay constant} 

We may recall firstly that the decay constant is inherited from the PQ symmetry-breaking vacuum expectation value.
Above the scale of `ordinary' PQ breaking, and obviously below that of monopole formation, we can expect monopole effects to be the only source of PQ symmetry breaking.

This being the case, the axion decay `constant' should again be a purely geometric factor in a single monopole background, and more generally proportional to the monopole number density in the environment of a monopole/anti-monopole plasma.
Although irrelevant for our present-day Universe, this is an intriguing point in that at appropriate monopole densities axion interactions could be essentially unsuppressed.

Once the mechanism of `ordinary' PQ breaking sets in, whichever symmetry-breaking effect is larger will dominate, and at low energies and/or monopole densities the usual PQ scale will prevail.

\subsection{The Peccei-Quinn mechanism}

The PQ mechanism requires a complex scalar $S$ with a symmetry breaking potential
\begin{align}
	\mathcal{L}_{PQ}=|\partial_\mu S|^2-\lambda_S\left(|S|^2-f^2\right)^2\,.
\end{align}
For simplicity we focus on the KSVZ (Kim-Shifman-Vainshtein-Zakharov) case \cite{Kim:1979if, Shifman:1979if}, where heavy quarks are added via
\begin{align}
	\mathcal{L}_Q=-\lambda_QS\overline Q_RQ_L+h.c.\,,
	\label{KSVZ interaction}
\end{align}
such that there exists an additional symmetry, under
\begin{align}
	S\to e^{2i\alpha} S\,,\quad
	Q\to e^{i\alpha\gamma_5}Q\,.
\end{align}
Once this is broken the axion appears as the associated Goldstone mode, identified as the phase of $S$ about $\left\langle S\right\rangle$.

Ordinarily the next step to deriving the axion potential is to expand about $\left\langle S\right\rangle$ and replace the fermion bilinears present with their vacuum expectation values, as is appropriate below the scale of confinement.
In the case of the QCD axion this potential then minimises $\left\langle\overline\theta\right\rangle$, thereby solving the strong CP problem.

However, as demonstrated in the investigations of the Callan-Rubakov effect, in a one-monopole background fermion bilinears such as $\left\langle\overline Q_LQ_R\right\rangle$ gain a non-zero value proportional simply to the monopole proximity \cite{Rubakov:1982fp}
\footnote{Strictly speaking they are $\sim r^{-3}$ for $r\lesssim 1/m$, and exponentially decay at distances beyond this \cite{Nair:1983ps}.}.
In fact it is precisely bilinears of the form $\left\langle\overline p_Le_R\right\rangle\sim r^{-3}$ which indicate the possibility of monopole-induced proton decay into a positron.
Of course, since monopoles couple with the inverse of the usual gauge coupling, strong coupling phenomena such as fermion condensation should not be unexpected.
Any bilinears which gain a vacuum expectation value in this way and carry PQ charge will break the associated symmetry.

We can then construct the effective potential for $S$ in a monopole background by replacing these bilinears by their monopole-induced vacuum expectation values.

\subsection{Monopole induced effective potential}

At temperatures high enough to negate `ordinary' PQ symmetry breaking, we have the effective potential
\begin{align}
	V(S)
	=\left(\lambda_S|S|S+m_S^2S-\lambda_Q\left\langle\overline Q_LQ_R\right\rangle\right)S^\dagger+h.c.\,,
\end{align}
where $\left\langle S\right\rangle\simeq\lambda_Q\left\langle\overline Q_LQ_R\right\rangle/2m_S^2+\mathcal{O}(\lambda_S)$, and $m_S^2\propto T^2$ is the thermal mass. 
In the early Universe we can define an effective average value for bilinears like this via the typical inter-monopole separation, given by the monopole number density $n_M$.

Of course, this quantity is not constant in time.
Neglecting monopole-antimonopole annihilation and assuming the expansion of the Universe to be adiabatic, we have
\begin{align}
	\left\langle\overline Q_LQ_R\right\rangle\sim n_M\propto T^3\,.
\end{align}
Actually, a little more care is required here in that computing this quantity in the presence of a $\theta$-angle
\footnote{There is also a subtlety here in that we may encounter a distinct $\theta$ angle for each gauge group present.
As explored in \cite{Simic:1986pd}, monopoles might locally enforce $\left\langle\theta_{EM}\right\rangle\simeq0$ at the expense of $\left\langle\theta_{QCD}\right\rangle\neq0$. 
Generally speaking, by $\theta$ here we mean $\theta_{QCD}$.}, cluster decomposition implies $\left\langle\overline Q_LQ_R\right\rangle\sim e^{i\theta}n_M$ \cite{Shnir:2005xx}.

Expanding \eqref{KSVZ interaction} about $\left\langle S\right\rangle$ we find the axion potential
\begin{align}
	V(a)
	\simeq\lambda_Q\left\langle S\right\rangle n_M\left((1-\cos\left(\frac{a}{\left\langle S\right\rangle}+\theta\right)\right)\,,
\end{align}
where the sine term has been cancelled by the Hermitian conjugate, and a constant shift has been included to reflect the typical energetic preference for $\left\langle a/S\right\rangle+\theta=0$.

Since the `heavy' $Q_{L/R}$ quark mass derives from a Yukawa coupling to $\left\langle S\right\rangle$, which will decrease with monopole dilution, we needn't worry about entering a regime where finite mass effects leave $\left\langle\overline Q_LQ_R\right\rangle$ exponentially suppressed.

However at low temperatures we expect conventional PQ symmetry breaking to also occur, such that monopole-induced breaking effects may be subleading.
In this phase we can minimise to find $\left\langle S\right\rangle$ as before, but it is simpler to see by eye the effects of the monopole background, as in Figure 1.
\begin{figure}[h!]
	\centering
	\includegraphics[width=0.6\linewidth]{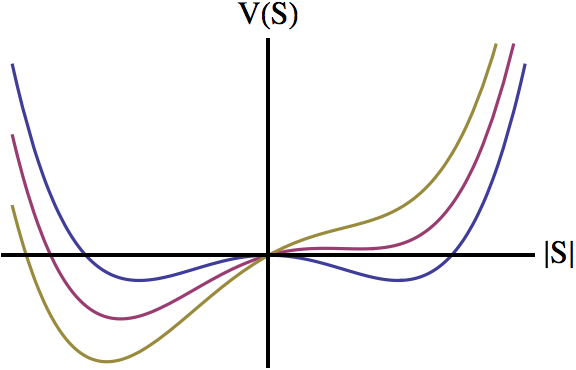}
  	\caption{The potential for the PQ scalar after `ordinary' PQ symmetry breaking has taken place, for different values of $\left\langle\overline Q_LQ_R\right\rangle$.
	Moving closer to the monopole (or increasing the monopole density) increases $\left\langle\overline Q_LQ_R\right\rangle$ and thereby the tilt of the potential, leading to a small modification of $\left\langle S\right\rangle$.
Since the axion arises from the angular component of $S$, this also leads to the usual picture of the axion potential being tilted via anomalous effects to generate a non-zero mass, although in this instance, as expected, they are directly visible at leading order in the semiclassical expansion.}
\end{figure}

Therefore, prior to the QCD phase transition, but in the vicinity of the PQ phase transition $\left\langle S\right\rangle\sim f$ as the `ordinary' PQ symmetry breaking has become dominant due to dilution of the monopoles.
The axion decay constant is then $f$ as usual, whilst the mass arises purely due to monopole effects, giving $m_a^2\simeq \lambda_Qn_M/f$.

At this point the masses of the `heavy' quarks will be fixed to approximately their usual values as $\left\langle S\right\rangle\sim f$, and they can be integrated out.
This is transmitted in the usual way to the low energy theory via the axial anomaly so that now $m_a^2\simeq m_u n_M/f^2$, and once QCD confinement occurs the ordinary axion mass will also appear, and subsequently dominate at low monopole densities.

However, we should be careful in that these terms are not the only contributions to $\partial_\mu J_5^\mu$, and hence the axion potential.
In fact, as we will see they should be subleading in our regime of interest.
From \eqref{axial anomaly} we firstly expect a contribution of the form $Mn_M$, which enters via the current algebra equivalence $\partial_\mu J_5^\mu\sim m_a^2f^2$.

For the minimal Julia-Zee dyon in a grand unified theory, $\int\partial_\mu J^\mu_5=m_X/\pi$ \cite{Marciano:1976as}, where $m_X$ is the superheavy gauge boson mass
\footnote{There is an implicit overall factor of $N$ here, corresponding to the number of PQ-charged fermions.}. 
Then by virtue of \eqref{axial anomaly} and the relation from the previous paragraph,
\begin{align}
	m_a^2\simeq m_{X}n_M/\pi f^2\,.
	\label{post PQ axion mass}
\end{align}
Gratifyingly, this expression is rather similar to the result derived in \cite{Fischler:1983sc} by entirely different means,
\begin{align}
	m_a^2\simeq e^2m_Xn_M/64\pi^3f^2\,.
\end{align}

The relative difference between these expressions can be traced back to the use in \cite{Fischler:1983sc} of an explicit cutoff at the radius of the monopole core $r_c\sim1/m_X$.
By also incorporating the contribution of the core, where most of the mass is concentrated, we gain a factor of $64\pi^2/e^2$.
At the GUT scale, this is an enhancement of $\mathcal{O}(10^{3})$.

The corresponding loss of coupling constant dependence should be also expected, since the axial charge remains quantised even when the corresponding symmetry is violated, and so in principle $\partial_\mu J^\mu_5\in\mathbb{Z}$.

Furthermore, we recall that these prior results were purely reliant upon electromagnetic effects, and as such strongly diminished by the screening effect of light fermions.
Whilst we also expect electromagnetic screening, prior to the QCD phase transition GUT monopoles will also carry unconfined colour charge which will in contrast be anti-screened. 
As such these findings apply not only for hidden sector monopoles, as in \cite{Kawasaki:2017xwt,Kawasaki:2015lpf,Nomura:2015xil}, but also to those from the visible sector too.

Although more involved it is straightforward to repeat these arguments for the DFSZ (Dine-Fischler-Srednicki-Zhitnitsky) case \cite{Dine:1981rt}.
Therein we make use of two Higgs doublets, with couplings of the form
\begin{align}
	\mathcal{L}=\lambda_u\, \overline q_LH_u u_R +\lambda_d\, \overline q_L H_d d_R -\lambda_HS^2H_u H_d\,,
\end{align}
and subsequent invariance under 
\begin{align}
	H_{u,d}\to e^{2i\alpha}H_{u,d}\,,\quad
	S\to e^{-2i\alpha}S\,,\quad 
	q\to e^{i\alpha\gamma_5}q\,.
	\label{DFSZ transformations}
\end{align}
The axion in this case will then be a linear combination of the phases of these three scalars.

An interesting corollary of this result is that in addition to PQ symmetry, electroweak symmetry can also be broken in the presence of monopoles.
We should note however that although more obvious, this is not unique to the DFSZ axion scenario. 
Generic bilinears which are charged under electroweak symmetry, such as $\left\langle \overline u_Lu_R\right\rangle$, will regardless be non-zero in the presence of a suitable GUT monopole.

\section{Axion abundance and isocurvature suppression}

These points established, we can now assess their role in the early Universe.
As previously noted, whilst the PQ mechanism offers the most elegant solution to the strong CP problem, the presence of such a light, long lived scalar carries a number of cosmological consequences.
Temporarily neglecting monopole effects, we can summarise these as follows.

\subsection{PQ symmetry unbroken during inflation}

If firstly $f$ is below the energy scale of inflation, PQ symmetry will be unbroken during that epoch.
Once symmetry breaking does occur the axion field in each Hubble patch will take a random initial value $\theta_i$, with an average background value of $\left\langle\theta_i\right\rangle=\pi^2/3$.
After the QCD phase transition, these initial misalignments lead to coherent oscillations, which contribute to the axion dark matter abundance.

Furthermore, we must contend with the possibility of axionic domain walls and strings, undiluted by inflationary expansion.
If stable they will quickly dominate the energy density of the Universe, whilst if unstable their decays will also contribute to the dark matter abundance.

For a domain wall number of unity, these effects combined can lead to
\begin{align}
	\Omega_ah^2\simeq 4\left(\frac{f}{10^{12}\mathrm{GeV}{}^{-1}}\right)^{1.19}\,,
	\label{PQ unbroken axion abundance}
\end{align}
where $h$ is the current-day dimensionless Hubble parameter, and we neglect anharmonicities in the axion potential \cite{Kawasaki:2014sqa}.
Evidently this is in tension with the observed dark matter abundance $\Omega_{DM}h^2\simeq0.12$, unless $f$ is sufficiently low.

\subsection{PQ symmetry broken during inflation}

If $f$ exceeds the energy scale of inflation, our visible Universe will comprise a patch taking a single value of $\theta_i$.
From \cite{Bae:2008ue}, the dark matter abundance due to the QCD axion is then
\begin{align}
	\Omega_ah^2\simeq 0.2\,\theta_i^2\left(\frac{f}{10^{12}\mathrm{GeV}{}^{-1}}\right)^{1.19}\,.
	\label{PQ broken axion abundance}
\end{align}
Whilst observational constraints can be accommodated for $f\sim10^{12}$ GeV, if $f>>10^{12}$ GeV, as is favoured by string theory, $\theta_i$ must be finely tuned close to zero.

Furthermore, we must account for the behaviour of the axion during inflation, where sufficiently light fields will accumulate quantum fluctuations, with amplitude $\delta\theta=H_I/2\pi$.
Once the axion becomes massive, these fluctuations can be converted into isocurvature density perturbations in the early Universe, which in turn lead to anisotropies in the CMB.

Results from the Planck mission constrain the isocurvature fraction to be 
\begin{align}
	\frac{\Omega_a}{\Omega_{DM}}\frac{\delta\theta\left(T_{QCD}\right)}{\theta_i}\lesssim 4.8\times 10^{-6}\,,
	\label{isocurvature fraction}
\end{align}
where $\delta\theta\left(T_{QCD}\right)$ is the average axion angular fluctuation at $T_{QCD}\sim 1$ GeV \cite{Ade:2015lrj,Choi:2015zra}.

Assuming a standard cosmological history, this can be used to infer that
\begin{align}
	H_I\lesssim9\times10^{8}\mathrm{GeV}\sqrt{\frac{\Omega_{DM}}{\Omega_a}}\left(\frac{f_a}{10^{16}\mathrm{GeV}{}^{-1}}\right)^{0.4}\,.
\end{align}
This is typically taken to imply the incompatibility of $f_a>> 10^{11}$ GeV for the QCD axion and an observable tensor to scalar ratio. 

\subsection{Monopole effects}

In this context, the mass created by a time-dependent population of monopoles is a significant benefit.
Indeed, a time-dependent mass is exactly the solution to the cosmological problems created by light long-lived scalars suggested in \cite{Linde:1996cx}, as this allows these fields to be pushed toward and then trapped in their minima, to evolve from there almost adiabatically toward their low-energy equilibria.

By then dynamically reducing $\left\langle\theta_i\right\rangle$ and $\left\langle\delta\theta\right\rangle$ at the time of the QCD phase transition, axion-sourced isocurvature and the contribution of the misalignment mechanism to the axion dark matter abundance can be safely reduced.
Because the modified axion potential will evolve adiabatically with the expansion of the Universe, $\left\langle a\right\rangle$ will also ultimately take the value required to cancel the strong CP phase \cite{Kawasaki:2015lpf}.

Our quantity of interest in this regard is $T_{osc}$, the temperature at which $m_a(T_{osc})\simeq H(T_{osc})$ and the axion will begin to oscillate.
We can solve for this via the Friedmann equation $3H^2M_P^2=\pi^2g^*T^4/30$, where $g^*$ is the number of relativistic degrees of freedom at that temperature.

Since cosmological problems primarily appear for large $f$, we assume PQ symmetry is broken prior to monopole formation and use \eqref{post PQ axion mass}.
The Friedmann equation gives 
\begin{align}
	T_{osc}\simeq 750\, \mathrm{GeV}\left(\frac{Y_M}{10^{-26}}\right)\left(\frac{10^{12} \mathrm{GeV}}{f}\right)^{2}\,,
\end{align}
where we set the unification scale to $10^{16}$ GeV, so that $m_{mono}\sim 10^{17}$ GeV, and $Y_M\equiv n_M/s$ is the monopole number per comoving volume, with $s$ the entropy density.
This then allows the the onset of axion oscillations much earlier in cosmological history.

Since the observational bound on $Y_M$ for superheavy monopoles is derived from the survival of galactic magnetic fields, we note that it may actually underestimate the monopole number density during our epoch of interest.
For example, GUT monopoles may form which are uncharged under $U(1)_{EM}$ and so are unstable after colour confinement\cite{Lazarides:1980va}.
These will contribute to the early-time axion mass, but decay thereafter.

However, for simplicity we will assume that the standard bounds are valid and $Y_M\lesssim 10^{-26}$ prior to the QCD phase transition \cite{Parker:1970xv}.
Although recent arguments suggest that due to the astrophysical behaviour of monopole plasma this bound could be further reduced by $\mathcal{O}(10^2)$ \cite{Medvedev:2017jdn}, we can in that case still find $T_{osc}>>T_{QCD}$.

From \cite{Kawasaki:2015lpf} we have $\Omega_ah^2\simeq3\times10^{-14}\theta_i^2f/T_{osc}$ in this scenario, using $g^*=106.75$, which becomes
\begin{align}
	\Omega_ah^2\simeq4\times10^{-5}\theta_i^2\left(\frac{10^{-26}}{Y_M}\right)\left(\frac{f}{10^{12} \mathrm{GeV}}\right)^{3}\,,
	\label{axion abundance}
\end{align}
where we can estimate the `PQ unbroken' case by substituting the misalignment angle dependence for an additional overall factor of $\sim 20$, following \eqref{PQ unbroken axion abundance} and \eqref{PQ broken axion abundance}.

It is straightforward to see that in either instance a significant reduction of the axion dark matter abundance is possible.
For large $f$ equation \eqref{axion abundance} may appear to predict more dark matter than the standard scenario, but this is because $T_{osc}<T_{QCD}$ in that case and monopole effects are negligible.

From \eqref{isocurvature fraction} we can see that suppressing axion dark matter will reduce the corresponding isocurvature contribution.
However we should also account for the reduction in the angular fluctuation $\delta\theta(T_{QCD})$, which satisfies $\delta\theta(T)\simeq\delta\theta(T_{osc})(T/T_{osc})^{3/4}$ \cite{Nomura:2015xil}.

To assess the phenomenological implications of both these effects, we can for concreteness focus on the `PQ broken' scenario and assume $Y_M\simeq10^{-26}$.
The resultant parameter space is given in Figure 2, where we neglect anharmonic effects in the axion potential, but include the back-reaction of isocurvature on the axion misalignment population \cite{Marsh:2015xka}. 

\begin{figure}[h!]
	\centering
	\includegraphics[width=0.95\linewidth]{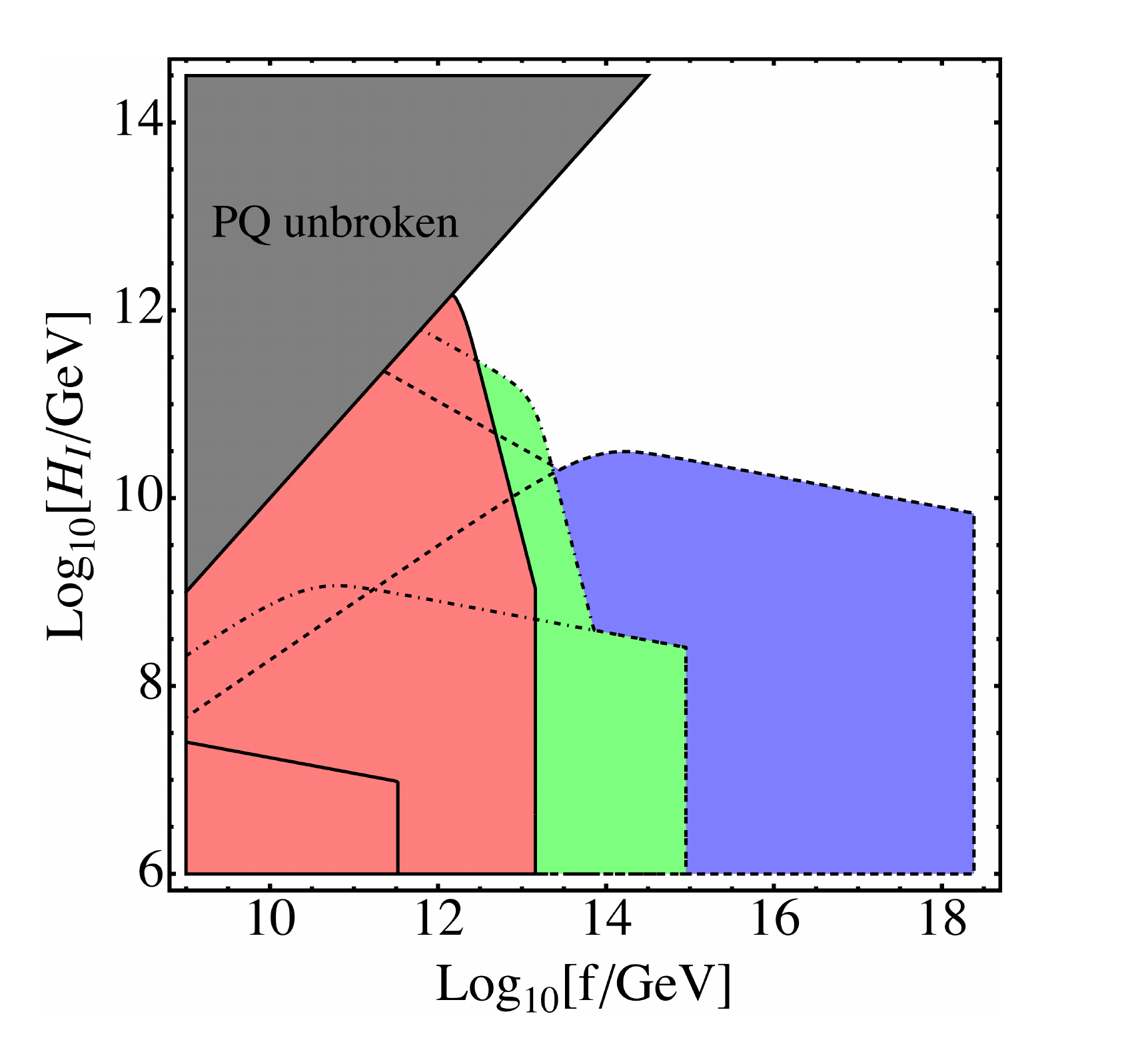}
  	\caption{The QCD axion parameter space satisfying both dark matter and isocurvature constraints in the `PQ broken scenario'.
	The (solid, dot-dashed, dashed) lines correspond to $\theta_i$ = $(1,10^{-2},10^{-4})$.
	In each case there is an upper and lower curve, corresponding to the allowed region with and without GUT monopoles satisfying $Y_M\simeq10^{-26}$.
	These upper and lower curves merge once $T_{osc}<T_{QCD}$, since then monopole effects are negligible.
	Colour is added as a visual aid.}
\end{figure}

As can be seen, the effect of GUT monopoles is most pronounced for $f\lesssim 10^{14}$ GeV, corresponding to when $T_{osc}\gtrsim T_{QCD}$, and the available parameter space there can be significantly enlarged.
In that regime the deviation from the monopole-free case arises primarily from the weakening of the isocurvature constraint, since both $\Omega_a$ and $\delta\theta(T_{QCD})$ are reduced
\footnote{The $T_{osc}$ dependence of $\Omega_a$ and $\delta\theta(T_{QCD})$ obviously differs, and so we should account for the suppression of both these quantities in computing the isocurvature contribution.}.

\section{Conclusions \& discussion}

In this letter we have examined in the unique interactions of axions and monopoles induced by the Witten effect, building upon a number of previous studies.

By generalising these efforts to the full QFT context we have hopefully better illuminated the ability of monopoles to break PQ symmetries directly, and induce an axion potential free of the usual instanton-related exponential suppression factor.
As a result we have found the monopole-induced mass in this context to be significantly enhanced relative to the corresponding value found in prior studies, which took place in the simpler context of axion electrodynamics.

This time-dependent axion mass is particularly useful in the early Universe, in that it allows the axion field to begin oscillating well before the QCD phase transition.
This can in turn reduce the overproduction of axionic dark matter and isocurvature perturbations in the CMB, which are particularly harmful for large $H_I$ and $f$.
As a result of the enhancement we have demonstrated here, this can also be achieved with a phenomenologically acceptable population of GUT monopoles, rather than requiring the use of a hidden sector.

We note that the experimental discovery of the QCD axion could then lend support to the existence of magnetic monopoles, if there is a mismatch between the implied and observed abundance and isocurvature contribution, suggestive of some suppression mechanism.

These findings may also have wider applicability in the context of string-theoretic axions, and the associated cosmological moduli problem.

\section*{Acknowledgments}

This research was supported in part by a CAS President's International Fellowship, the Projects 11475238 and 11647601 supported by the National Natural Science Foundation of China, and by the Key Research Program of Frontier Science, CAS.

\end{document}